# Gravitational Poynting Vector and Gravitational Larmor Theorem in Rotating Bodies with Angular Acceleration

C. J. de Matos[*]

*ESA-ESTEC, Directorate of Scientific Programmes, PO Box 299, NL-2200 AG Noordwijk, The Netherlands*

M. Tajmar[†]

*Austrian Research Centers Seibersdorf, A-2444 Seibersdorf, Austria*

**Abstract**

The gravitational Poynting vector, $\vec{S}_g = \frac{c^2}{4\pi G}\vec{\gamma} \times \vec{B}_g$, provides a mechanism for the transfer of gravitational energy to a system of falling objects. In the following we will show that the gravitational poynting vector together with the gravitational Larmor theorem also provides a mechanism to explain how massive bodies acquire rotational kinetic energy when external mechanical forces are applied on them.

[*] Staff Member, Science Management Communication Division, Phone: +31-71-565 3460, Fax: +31-71-565 4101, E-mail: clovis.de.matos@esa.int

[†] Research Scientist, Space Propulsion, Phone: +43-50550-3142, Fax: +43-50550-3366, E-mail: martin.tajmar@arcs.ac.at

**Free Fall and Gravitational Poynting Vector**

Consider a cylinder of length $\ell$, radius $a$, mass $\rho$ per unit volume, moving downward with the influence of a gravitational field $\vec{\gamma}$.

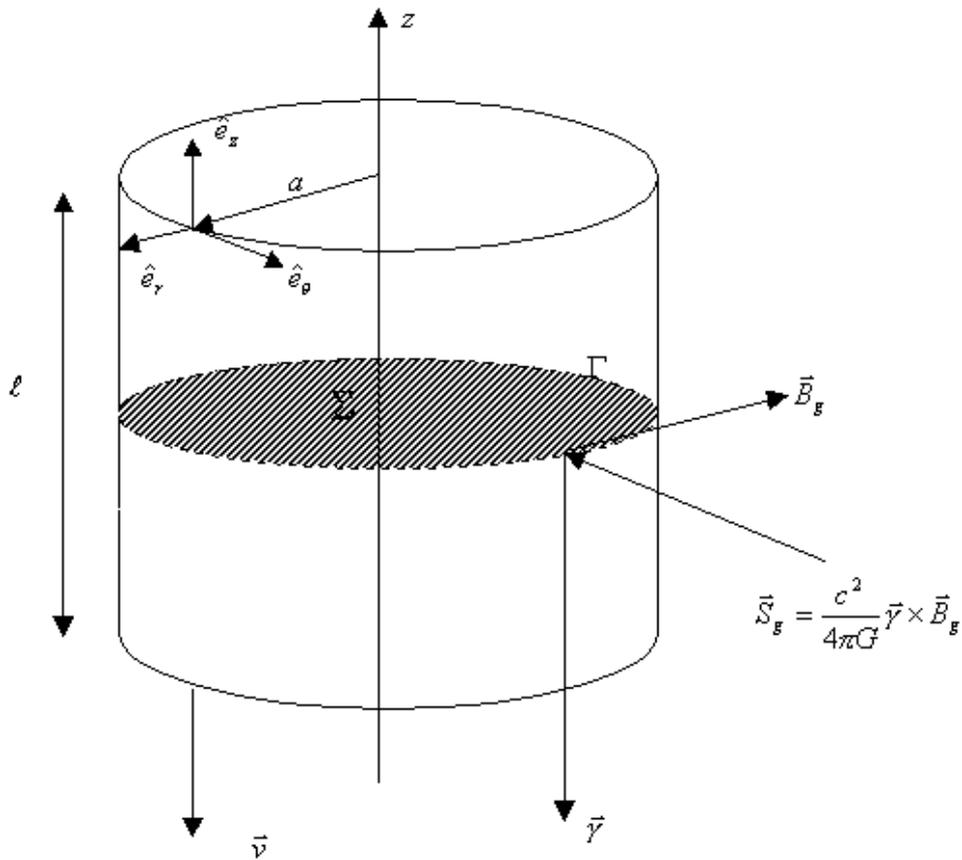

**Figure 1 Non-rotating cylinder falling in a gravitational field**

Let us use cylindrical coordinates for describing the motion of the cyclinder and the corresponding energy relations. The gravitational field around the cylinder is equal to the acceleration of Earth gravity $\vec{\gamma}$ and is directed downwards:

$$\vec{\gamma} = -\gamma \hat{e}_z \qquad (1)$$

Let the velocity of the cylinder be $\vec{v}$. The cylinder constitutes a mass current which is given by:

$$\vec{j} = \rho \vec{v} = -\frac{m\vec{v}}{\pi a^2 \ell} \hat{e}_z. \quad (2)$$

The gravitomagnetic field at the surface of the cylinder will be:

$$\text{Rot } \vec{B}_g = -\frac{4\pi G}{c^2} \vec{j}$$

$$\oiint_\Sigma \text{Rot } \vec{B}_g \cdot d\vec{\sigma} = -\frac{4\pi G}{c^2} \oiint_\Sigma \vec{j} \cdot d\vec{\sigma}$$

$$\oint_\Gamma \vec{B}_g \cdot d\vec{s} = -\frac{4\pi G}{c^2} \oiint_\Sigma \vec{j} \cdot d\vec{\sigma} \quad (3)$$

$$2\pi a B_g = -\frac{4\pi G}{c^2} j \pi a^2 \quad (4)$$

Inserting equation (2) into equation (4)

$$B_g = -\frac{2G}{c^2} \frac{mv}{\ell a} \quad (5)$$

The minus sign appearing in equation (5) is due to the fact that the direction of $B_g$ is given by the "left hand rule" since, contrary to electric case, like masses attract, therefore in the given situation represented in figure 1

$$\vec{B}_g = \frac{2G}{c^2} \frac{mv}{\ell a} \hat{e}_\theta \quad (6)$$

The gravitational Poynting vector at the cylinder surface is:

$$\vec{S}_g = \frac{c^2}{4\pi G} \vec{\gamma} \times \vec{B}_g = \frac{mv\gamma}{2\pi a \ell} \hat{n}_{in} \quad (7)$$

where $\hat{n}_{in} = -\hat{e}_r$ is a unit vector normal to the cylinder's surface and directed into the cylinder. Multiplying $\vec{S}_g$ by the area of the cylinder, we obain for the rate of gravitational energy influx into the cylinder:

$$\frac{dU}{dt} = \frac{mv\gamma}{2\pi a \ell} 2\pi a \ell = mv\gamma \quad (8)$$

The rate at which the kinetic energy of the cylinder increases is:

$$\frac{dU}{dt} = \frac{d}{dt}\left(\frac{mv^2}{2}\right) = mv\frac{dv}{dt} = mv\dot{v} \qquad (9)$$

We just have shown that the rate at which the kinetic energy of the cylinder increases is completely accounted by the influx of gravitational energy into the cylinder.

**The principle of equivalence in the case of reference frames with angular acceleration and the gravitational Larmor theorem**

The gravitational Larmor theorem (GLT) states that a gravitomagnetic field is locally equivalent to a rotating reference frame with angular velocity

$$\vec{B}_g = -\vec{\Omega} \qquad (10)$$

The GLT extends the classical principle of equivalence (which contend only linearly accelerated reference frames, $\vec{\gamma} = -\dot{\vec{v}}$) to rotating reference frames. Einstein's heuristic principle of equivalence is traditionally stated in terms of the translational acceleration of the "Einstein elevator"; however, it is clear from the analysis of this section that a larmor rotation of the elevator is necessary as well to take due account of the gravitomagnetic field.

A time varying angular velocity flux will therefore be associated with a non-Newtonian gravitational field proportional to the tangential acceleration. The gravitational electromotive force produced in a gyrogravitomagnetic experiment can be calculated using the gravitational Faraday induction law:

$$\varepsilon_g = \oint_\Gamma \vec{\gamma} \cdot d\vec{l} = -\frac{d\phi_{gm}}{dt} = -\frac{d}{dt}\iint_\Sigma \vec{B}_g \cdot d\vec{\sigma} \qquad (12)$$

(where $\varepsilon_g$ [J / Kg] is the gravitational electromotive force) together with the GLT expressed through equation (10)

$$\varepsilon_g = \frac{d}{dt}\iint_\Sigma \vec{\Omega} \cdot d\vec{\sigma} \qquad (13)$$

The induced non-Newtonian gravitational field associated with this gravitational electromotive force is:

$$\oint_\Gamma \vec{\gamma} \cdot d\vec{l} = \frac{d}{dt} \iint_\Sigma \vec{\Omega} \cdot d\vec{\sigma} \qquad (14)$$

**Energy transfer to spinning masses with angular acceleration**

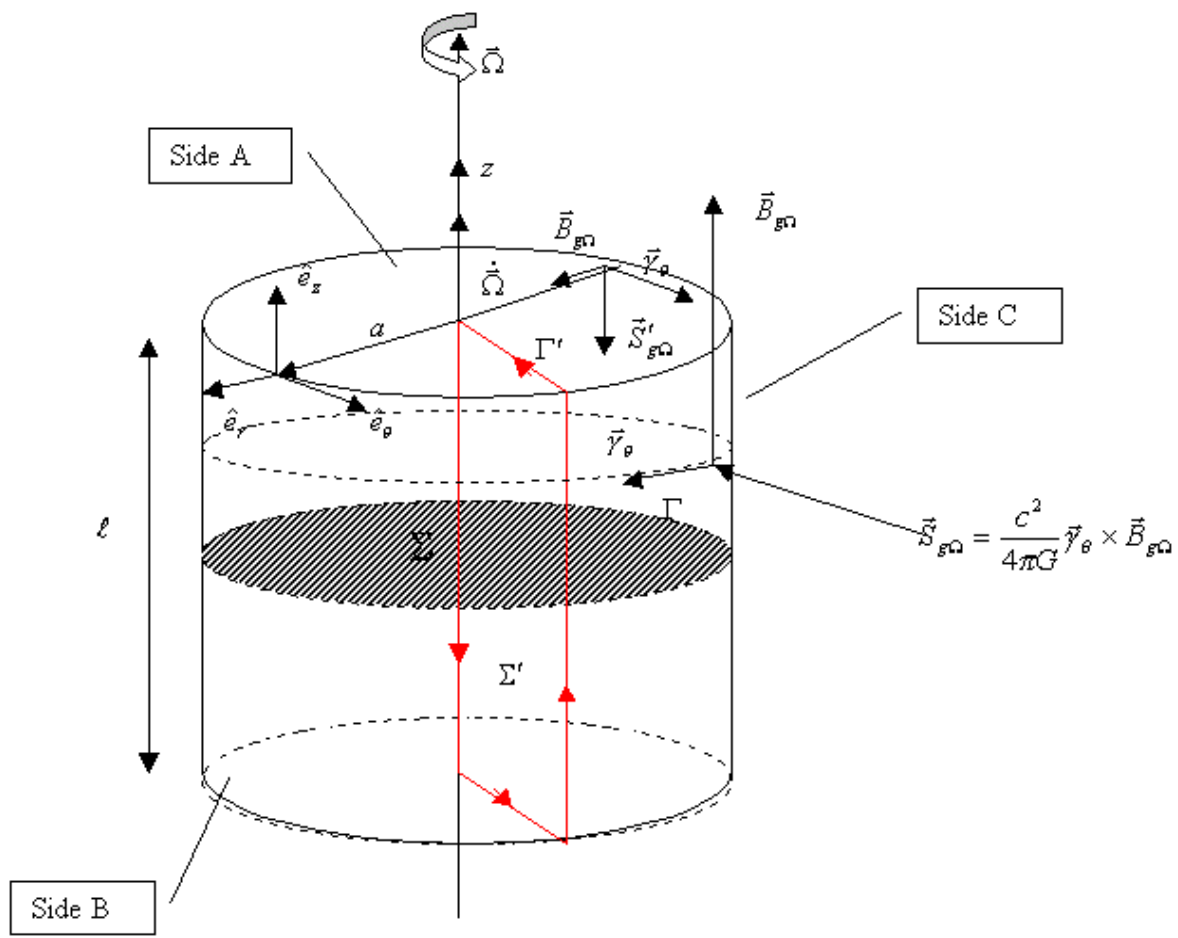

**Figure 2 cylinder with angular acceleration, up side-A, down side-B, cylindrical side-C**

Let us compute the mass current through section $\Sigma'$.

$$d\vec{j}_\Omega = \rho\, d\vec{v}$$

$$d\vec{j}_\Omega = \rho\, \Omega\, dr\, \hat{e}_\theta$$

$$\int d\vec{j}_\Omega = \int_{r=0}^{r=a} \rho\, \Omega\, dr\; \hat{e}_\theta$$

$$\vec{j}_\Omega = \rho\, \Omega\, a\, \hat{e}_\theta \qquad (16)$$

The gravitomagnetic field at the surface of the cylinder due to the mass current generated by the rotation of the cylinder around its longitudinal axis is:

$$\oint_{\Gamma'} \vec{B}_{g\Omega} \cdot d\vec{s} = -\frac{4\pi G}{c^2} \iint_{\Sigma'} \vec{j}_\Omega \cdot d\vec{\Sigma} \qquad (17)$$

Doing (16) into (17)

$$B_{g\Omega}(2a + 2\ell) = \frac{4\pi G}{c^2} \rho\, \Omega a\, a\ell$$

$$B_{g\Omega} = \frac{2Gm\Omega}{c^2} \frac{1}{a+\ell} \qquad (18)$$

From (14) we can compute the induced non-Newtonian gravitational field to which is submitted the cylinder due to its angular acceleration:

$$\oint_\Gamma \vec{\gamma}_\theta \cdot d\vec{l} = \frac{d}{dt} \iint_\Sigma \vec{\Omega} \cdot d\vec{\sigma}$$

$$2\pi a\, \gamma_\theta = \dot{\Omega}\, \pi a^2$$

$$\vec{\gamma}_\theta = \frac{1}{2} a\dot{\Omega}\, \hat{e}_\theta \qquad (19)$$

From (18) and (19) we compute the gravitational Poynting vector for side C

$$\vec{S}_{g\Omega} = \frac{c^2}{4\pi G} \vec{\gamma}_\theta \times \vec{B}_{g\Omega} = \frac{1}{4\pi}\left(\frac{a}{a+\ell}\right) \Omega\dot{\Omega} m\, \hat{n} \qquad (20\text{-C})$$

$\hat{n}$ will be equal to $+\hat{e}_r$ or $-\hat{e}_r$ depending on the sign of the angular acceleration $\dot{\Omega}$. For sides A and B we will have:

$$\vec{S}'_{g\Omega} = \frac{c^2}{4\pi G} \vec{\gamma}_\theta \times \vec{B}_{g\Omega} = \frac{1}{4\pi}\left(\frac{a}{a+\ell}\right) \Omega\dot{\Omega} m\, \hat{n}' \qquad (20\text{-A,B})$$

$\hat{n}'$ will be equal to $+\hat{e}_z$ or $-\hat{e}_z$ depending on the sign of the angular acceleration $\dot{\Omega}$. From (20-C) and (20-A,B) we deduce that the rate at which the mechanical energy is absorbed by sides A, B, and C is given by:

$$\left.\frac{dU}{dt}\right|_\Omega = S'_{g\Omega} 2\pi a^2 + S_{g\Omega} 2\pi a \ell$$

$$\left.\frac{dU}{dt}\right|_\Omega = \frac{1}{2} ma^2 \Omega \dot{\Omega} \quad (21)$$

The rate at which the rotational kinetic energy of the cylinder increases is

$$\left.\frac{dU}{dt}\right|_{T\Omega} = \frac{d}{dt}\left(\frac{1}{2} I\Omega^2\right) = I\Omega\dot{\Omega} = \frac{1}{2} ma^2 \Omega \dot{\Omega} \quad (22)$$

$I = \frac{1}{2} ma^2$ being the moment of inertia of the cylinder.

**Conclusion**

Comparing equations (21) and (22) we see that the rate at which the rotational kinetic energy of the cylinder increases (or decreases) due to the application of external mechanical forces on the cylinder, is completely accounted by the influx (out-flux) of gravitational energy into (outward) the cylinder. This demonstrates the validity of the gravitational Larmor theorem, and shows how the transfer of mechanical work to a body can be interpreted as a flux of gravitational energy associated with non-Newtonian gravitational fields produced by time varying angular velocities. This is an encouraging result regarding the possible detection of macroscopic non-Newtonian gravitational fields induced through the angular acceleration of the cylinder in the region located outside the rotating cylinder. The non-Newtonian gravitational field outside the cylinder is given by:

$$\gamma = \frac{1}{2} \frac{a^2}{r} \dot{\Omega} \quad (23)$$

where $r > a$ is the distance from the cylinder's longitudinal axis. For $r \leq a$ we have, $\gamma = \frac{1}{2}a\dot{\Omega}$. For the following values of $a = 0.1m$, $r = 1m$, $\dot{\Omega} = 200[Hz/s]$, $\gamma$ will have the value of $1[m/s^2]$. We recommend that experiments shall be performed with the aim of evaluating equation (23). Finally the experimental detection of this effect would be an indirect confirmation of the gravitomagnetic Barnett effect.